Mohsen Imani*, Armon Barton, and Matthew Wright

# Forming Guard Sets using AS Relationships

**Abstract:** The mechanism for picking guards in Tor suffers from security problems like *guard fingerprinting* and from performance issues. To address these issues, Hayes and Danezis proposed the use of *guard sets*, in which the Tor system groups all guards into sets, and each client picks one of these sets and uses its guards. Unfortunately, guard sets frequently need nodes added or they are broken up due to fluctuations in network bandwidth. In this paper, we first show that these breakups create opportunities for malicious guards to join many guard sets by merely tuning the bandwidth they make available to Tor, and this greatly increases the number of clients exposed to malicious guards. To address this problem, we propose a new method for forming guard sets based on Internet location. We construct a hierarchy that keeps clients and guards together more reliably and prevents guards from easily joining arbitrary guard sets. This approach also has the advantage of confining an attacker with access to limited locations on the Internet to a small number of guard sets. We simulate this guard set design using historical Tor data in the presence of both relay-level adversaries and network-level adversaries, and we find that our approach is good at confining the adversary into few guard sets and thus limiting the impact of attacks.

## 1 Introduction

Tor allows clients to create anonymous connections to their desired destinations via three-hop encrypted channels called *circuits*. A circuit is built over a path of three relays, an *entry*, a *middle*, and an *exit*, selected from among the thousands of volunteer relays distributed across the globe. In Tor, no single relay in the circuit nor any third party in the network should be able to link the source with the destination.

**\*Corresponding Author: Mohsen Imani:** The University of Texas at Arlington, E-mail: mohsen.imani@mavs.uta.edu
**Armon Barton:** The University of Texas at Arlington, E-mail: armon.barton@mavs.uta.edu
**Matthew Wright:** Rochester Institute of Technology, E-mail: mwright@rit.edu

Since relays are run by volunteers, however, it remains a risk that multiple relays on a circuit are run by a single entity who could then break the user's anonymity. In fact, if all relays on the circuit were picked at random every time, a Tor user would be rolling the dice with her privacy every few minutes. Most circuits would be fine, but eventually she would roll a pair of malicious relays and lose her anonymity. To prevent the majority of users from getting compromised, Tor fixes the client's entry node to be the same in every circuit for up to nine months. If this entry node, called a *guard*, is honest and does not get compromised, then the client's identity cannot be directly discovered by malicious relays while the guard is still being used [37].

A key design decision around the use of guards is how to assign guards to users. If a user picks a guard with very low bandwidth, as an extreme example, then not only will her performance be poor, she may be the only user regularly using that guard and can thus be profiled [19, 29, 31, 33]. This is known as *guard fingerprinting*. More generally, there are several anonymity and performance considerations for picking guards that have only recently been explored [8, 10, 13, 15].

One solution to the guard fingerprinting problem is to group all guards into *guard sets* [13] and have each client pick one of the guard sets and use the guards in this guard set for the first hop on all of its circuits. Hayes and Danezis [18] proposed the first guard set algorithm for use in Tor. This algorithm uses guard relays' bandwidth as the key criterion in forming guard sets, such that all sets have almost the same amount of bandwidth. They also presented techniques for maintaining the guard sets when there is churn.

**Contributions.** In this paper, we first demonstrate that the algorithms proposed by Hayes and Danezis have vulnerabilities that allow an attacker to compromise many guard sets in the presence of churn over time (§3). In particular, we describe attacks that leverage the fact that the attacker controls the amount of bandwidth it makes available for a given guard node. With these attacks, a low-bandwidth adversary controlling 1% of total guard bandwidth can infiltrate around 40% of all guard sets within four months, and a high-bandwidth adversary controlling around 25% of total guard bandwidth can infiltrate 90% of guard sets.

To address these issues, we next propose a new design (§4) for guard sets that uses location in the Internet topology as the basis for building a hierarchy on top of the sets in which sets are built and maintained using guards that are topologically close to each other in the Internet. This limits an attacker's ability to compromise guard sets beyond whatever Internet locations he has access to. While bandwidth can be easily manipulated, most attackers will have a limit on the possible Internet locations of their guards.

We evaluate the security of this approach against attackers who control a fraction of the guards, with varying resource levels, using one network in the Internet (§5.1.1). Against a single malicious guard, the compromise rate after one year of running the attack is 0.076% compared to 0.044% for the prior approach. Against an attacker who controls 10% of Tor's guard bandwidth, the compromise rate compared to the prior approach dropped from 53% to 10% after one year and less than half of the rate for the current Tor design (23%). Against a botnet adversary, which inherently has a presence in more AS locations, the compromise rate fell from 53% to 37% compared to the prior approach after one year. Moreover, the fraction of compromised targets in our approach dropped from 98% to 44% compared to the prior approach in a targeted attack scenario (§5.1.2).

We also evaluate our approach against attackers who control one *AS* in the Internet (§5.1.3). We find that our approach has very similar results as both Tor and the prior work. Additionally, we merge our guard set design with DeNASA [9], a recently proposed AS-aware path selection algorithm, and show that the rate of streams being vulnerable to attack drops 80%.

Beyond this, we evaluate a number of other aspects (§5.2), including the sizes of anonymity sets, the bandwidth distribution among guard sets, and network performance. We conclude with a discussion (§6) of deployment and other issues to be addressed in future work.

## 2 Background

In this section, we briefly overview the AS structure of the Internet and the Tor anonymity system, and we then discuss related work.

### 2.1 Autonomous Systems

Our approach makes use of the structure of the Internet topology, so we describe the necessary concepts here.

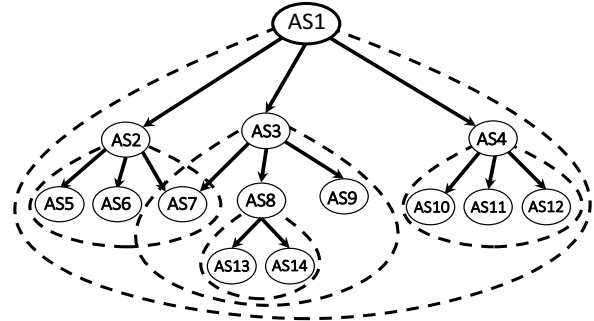

**Fig. 1. Customer cones**: The dashed lines show customer cones, the solid arrows are provider-to-customer links.

The network layer of the Internet is composed of *Autonomous Systems (ASes)* that are linked together by high bandwidth lines and fast routers. Each AS is owned and operated by one authority, such as a government, university, or Internet service provider. ASes contain a set of servers that are linked together in a LAN and are assigned IPs from an IP prefix that is unique to that particular AS. Relationships among ASes have been formed based on a variety of economic and political constraints. Using publicly available BGP table data, Gao [17] introduced a method that abstracts these relationships in three types: customer-to-provider (*c2p*), provider-to-customer (*p2c*), and peer-to-peer (*p2p*). In a c2p or p2c relationship, the customer provides monetary payment to the provider in exchange for the provider providing bandwidth to the customer. In a p2p relationship, the two ASes save monetary resources by exchanging traffic between one another on a quid-pro-quo basis.

Viewing ASes and their relationships in graph theoretic terms, we have a forest of trees, with backbone providers at the root nodes and customers as leaves. We can then define a *customer cone*, the set of ASes that can be reached from the root AS by only following the p2c links. For example, AS *A*'s customer cone consists of AS *A* plus AS *A*'s customers, plus AS *A*'s customers' customers and so on [27]. Figure 1 shows an example of how customer cones work, where AS1 has the biggest cone in the example and contains all the ASes shown, while AS8 has only AS13 and AS14 in its customer cone. The number of ASes in a customer cone and the number of unique IP prefixes advertised by these ASes are good metrics for ranking the size and importance of an AS [12]. Our guard set design groups the guard ASes that are in the same customer cone in the same set.

The underlying customer cone in our design can be built by different methods. The main requirement for our system are that the cones should be relatively stable over time compared with bandwidth fluctuations in the guard sets. We chose to use the *recursive customers* method [12, 27]. In this method, the customer cone of each AS is built by recursively visiting ASes reachable from that by p2c links. The customer cone of an AS is a subtree of customer ASes that can be reached from that AS. The size of a customer cone is the number of customer ASes in that cone. For the the customer cones in our experiments, we use the AS relationships provided by CAIDA for July 2015 [11].

## 2.2 Tor Overview

Tor is a volunteer-operated network that provides anonymity and privacy online. Tor has about 2 million daily users and 7,000 relays. Relays in Tor network, called *Onion Routers* (ORs), are run by volunteers who donate their bandwidth. ORs provide information about their donated bandwidth, IP address and ports, and *exit policies*—the addresses and ports they are willing to be connected to external Internet destinations—to a small group of servers called *directory authorities*. The directory authorities assign flags to some of the ORs based on their availability, bandwidth and exit policies and then they mutually agree upon a list called the *consensus* of all the information about all the ORs.

A Tor client, called the *Onion Proxy* (OP), first contacts one of directory authorities or their mirrors and downloads the consensus to get the current status of the Tor network. Since the Tor network is dynamic, with relays regularly joining and leaving, the directory authorities update the consensus hourly. The OP uses the consensus information to select a path of three relays to use in communicating with its destinations.

Once the OP picks this path of ORs, it then sets out to build a *circuit* of layered cryptographic connections through this path. Since relays in the circuits are selected from all the volunteer relays, it is possible that an adversary, who runs some guard or exit relays in the network, sits in the exit and entry position of some path. Such an adversary can observe the entry and exit traffic, correlate them, and link the client to her destinations [19, 24, 29, 35]. If the client chooses new relays for each circuit, she will eventually build a circuit in which the adversary's relays are in its entry and exit positions. To reduce the occurrence of this attack, Tor clients stick to using a single *guard* relay for the first relay on every path for months [13]. Guards should thus be stable, so that the client can rely on the guard to be available whenever it connects to Tor, and reasonably high bandwidth to prevent the guard from becoming a significant bottleneck to performance.

Directory servers keep track of relays' bandwidth and availability in the network and assign Guard flags to the relays that have the following criteria [8]:
– Have been continuously running for longer than 12.5% of the relays or at least eight days.
– Advertise bandwidth more than the median bandwidth of all the relays, or 2.0 MBps.
– Have a *weighted-fractional-uptime (WFU)*[1] of more than the median of all relays' WFU or 98% WFU.

## 2.3 Related Work

Elahi et al. [15] developed a framework called COGS to study guard selection schemes in Tor and evaluated the impact of churn, guard rotation, and the size of the guard list on clients' anonymity. Their results show that guard rotation exposes the users to more guards, increasing the chance of picking malicious guards. On the other hand, they find that guard rotation offers better load balancing on guards, better utilization of recently joined guards, and regaining the privacy of clients stuck using malicious guards. If Tor rotates the guards, they report, then larger guard lists lead to more compromises of anonymity; without guard rotation, larger guard lists lead to fewer compromises. They also find that larger guard lists lead to better, fairer performance.

Johnson et al. [24] evaluate the vulnerability of Tor to passive end-to-end correlation attacks from both relay- and network-level adversaries, with the settings used prior to 2014 of three guards rotated after 30 to 60 days. They define metrics that give us the probability of path compromise for a given user and the probability of time to the first compromise. They developed a path simulator that mimics the Tor client and implements multiple models of user activity, such as Web users and BitTorrent clients. They found that relay-level adversaries can maximize their resources by allocating more bandwidth to malicious guards than to malicious exits. Against an adversary running one 83.3 MBps guard re-lay and one 6.7 MBps exit relay, they find that 80% of users will be compromised within six months. Considering network-level adversaries, they find that an AS adversary can compromise 38% of Tor streams in three months, and a IXP adversary can compromise 20% of streams in the same period. We implement our guard selection scheme in their path simulator to analyze the security of our design at the network level.

---

[1] WFU is the percentage of the time that relay has been up, adjusted with a decay of 5% per every 12 hours the relay is off.

Based on the results from Elahi et al. and Johnson et al., Dingledine et al. [13] conclude that the guard selection mechanism in use prior to 2014 harmed users' security. Instead of using three guards for 30 to 60 days, they proposed using a single guard for nine to ten months. They note that, based on Elahi et al's findings, this proposal will provide stronger anonymity but suffers from poor performance and poor load balancing, as the newly joint guards will be underutilized. To fix these flaws, Dingledine et al. suggest raising the bandwidth bar in assigning guard flags from 250 KBps to 2MBps and having underutilized guards act as middle nodes. These changes were implemented in Tor and are still in effect as of the time of writing.

Dingledine el at. also suggested the idea of guard sets [13], which Hayes and Danezis [18] then expanded into a full proposal and evaluation. The Hayes and Danezis design puts the guards into sets based on their bandwidths, such that each set has approximately the same bandwidth. Although this approach is intuitive, we show in the following section how it leads to vulnerabilities that we then address in our proposed design.

## 3 Attacking Guard Sets

With guard sets, the directory authorities put all the guards into sets and include this assignment in the consensus [13]. The client randomly picks a guard set to use for a long period of time and then picks the guard for each circuit randomly from the selected set. The main advantage of this scheme is that it puts all the clients using a given guard set into one anonymity set, such that a guard fingerprinting attack could only identify one as a member of the set.

### 3.1 Hayes and Danezis Design

Hayes and Danezis performed the first detailed study of the guard set idea, and they propose algorithms for how to build guard sets, assign users to those sets, and maintain the sets as the Tor network changes [18]. To ensure load balancing, their proposal uses bandwidth as the main criteria to build the sets. In particular, it first uses the bandwidth values from the consensus to generate *bandwidth quanta*, where each quantum represents a block of bandwidth from a single guard node. Using an empirically selected threshold of 40 MBps, a guard's bandwidth is divided into multiple quanta such that each quantum is above the threshold. A guard that has bandwidth $BW$ generates $\lfloor \frac{BW}{40} \rfloor$ quanta, meaning that guards with less than 80 MBps bandwidth make up just one quantum. For example, if we have guards with bandwidths of 10, 70, and 90 MBps, we get quanta 10, 70, 45, and 45 MBs. The bandwidth quanta then are sorted from largest to smallest.

To build guard sets, the algorithm goes through the sorted list of quanta and moves one quantum at a time from the head of the list to the current set until the total bandwidth of the set reaches the threshold of 40 MBps. Then the current set is added to the list of sets, and a new set is started. If the leftover bandwidth quanta in the sorted list make up less than 40 MBps, they are not used to build sets and do not contribute to any guard sets. The goal of sorting the quanta list is to put guard nodes with similar bandwidth in the same set, and it also forces an attacker with many low-bandwidth guards into fewer sets with similar bandwidths instead of being spread out into sets with mixed bandwidths.

Over time, the total guard bandwidth in Tor fluctuates, as some new guards join the network and others go offline. These changes affect the bandwidth of guard sets and the available bandwidth quanta. To address this, the strategy of Hayes and Danezis is to first repair *damaged* guard sets with bandwidth of less than 20 MBps. When repairing a given damaged set, the algorithm finds the leftover bandwidth quanta that fall between between 50% to 100% of the maximum guard bandwidth of the set. This list of quanta is called the *candidate list* of the set. Quanta from the candidate list are added to the set one by one until the set's total bandwidth exceeds 40 MBps. Once all damaged guard sets are repaired, the algorithm builds a new guard set from any remaining leftover quanta if their combined bandwidth is more than 40 MBps.

#### 3.1.1 Vulnerabilities

Using bandwidth similarity to repair the sets opens a door for the attacker. The primary issue is that the attacker can identify guard sets that are close to break-

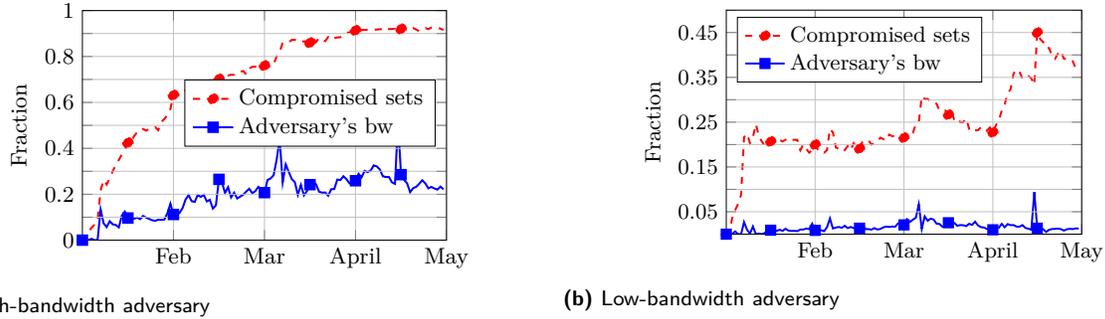

**(a)** High-bandwidth adversary

**(b)** Low-bandwidth adversary

**Fig. 2. Hayes-Danezis:** The fraction of compromised sets and fraction of the adversary's bandwidth to the total guard bandwidth

ing, i.e. around 20 MBps, and then add guards or tune his guards' bandwidths to have similar bandwidths. For example, if the vulnerable guard set has guards with bandwidths of 3–4 MBps, and the attacker has an unused guard with 5 MBps bandwidth, he can set it to offer a maximum of 3.5 MBps to Tor. Once these sets break, the repair algorithm will include the adversary's guards to be in the candidate list, increasing its chances of joining a particular set. Additionally, the attacker can create new compromised guard sets by adding his guards to the network or tuning their bandwidth so that the total bandwidth of leftover quanta and his guards is above 40 MBps, causing the algorithm to build a new set.

Another issue is that a guard set is not considered in need of repair if its bandwidth is at least 20 MBps. This means that as soon as the adversary joins a set, it can reduce the allocated guard bandwidth to the least possible value for being a guard, which can be as low as 2 MBps as long as the set's total bandwidth remains above 20 MBps. This can save the adversary's resources. Also, if an attacker gets one of his guards into a set, it will remain in that set forever even if all the other guards in the set are gone. This allows the attacker to retain a full guard sets' allocation of users while only using half of the bandwidth needed for building a new set.

## 3.2 Evaluation

We investigate the impact of these vulnerabilities on the adversary's ability to infiltrate guard sets and compromise Tor users. Jamie Hayes provided us with his implementation of the Hayes-Danezis algorithms for guard sets. We used their implementation and exploited the possible vulnerabilities in simulation.

### 3.2.1 Attacker Model

The goal of the attacker is to get into as many guard sets as possible and thereby compromise a large fraction of users. We run the Hayes-Danezis algorithm using consensus documents from January to May 2013, the same time period used by Hayes and Danezis for consistency. Like Hayes and Danezis [18], in all of our simulations we used the first consensus documents of each day, instead of hourly consensus updates, to allow for longer studies. We allow the attacker to add his guard relays to the network in the second day of simulation after the guard sets are formed, and the attacker aims to both get into new sets and into previously built sets.

We assume that the attacker has access to the guard relays' bandwidths, which he can obtain from the consensus and refine if necessary by periodic measurements. The attacker also keeps track of the assignment of guards to guard sets, which is also available in the consensus. The attacker uses this information to follow the sets' bandwidth and detect which sets are about to break. If a set is about to break and the set is already compromised, the attacker tunes his bandwidth to keep the set alive by keeping the set's bandwidth above 20 MBps. Otherwise, if a set is broken and not yet compromised, the attacker adds new guards to the network with bandwidths tuned to get added to the set's candidate list. Instead of adding new guard relays to the network, the attacker can reuse his unused guard relays and tune their bandwidth. If the attacker observes new sets are forming from leftover guards, the attacker adds new guards or re-uses his guards in the leftover quanta and tunes their bandwidth to get to the list of candidates for the new set. Because the quantum in the candidate list of a set are in descending order, The attacker only needs to tune his guard relay bandwidth slightly higher than the last bandwidth in the list which fixes the set.

In our simulations, we assume the attacker can add the new guards the same day he needs them. Note that the new sets are created and broken sets are repaired whenever there is enough bandwidth or a set is broken, but the changes are announced in the next consensus file. Also, the attacker's relays must be assigned the guard flag by the authorities. To do this, the attacker can run some relays that have all the criteria to get the guard flag (as listed in Section 2.2) except for one. For example, the attacker can run relays with high uptime but with bandwidth less than the minimum bandwidth required (currently 2 MBps), which saves his bandwidth while waiting for a set to break. When the attacker needs a new guard, he just needs to increase the bandwidth of the relay to get the Guard flag. Another technique would be for the attacker, who typically would run exit nodes to perform end-to-end correlation attacks together with his guards, to switch one of his exits to being a guard. This is easily done by first having a high-uptime relay with an exit policy, which will cause it to have the Exit-Guard flag, but it will be used exclusively as an exit due to exits being the bandwidth bottleneck in Tor. Then, to switch it to a guard, the attacker simply removes the exit policy.

We used two adversary models, a *high-bandwidth adversary* who controls about 25% of Tor's bandwidth and a *low-bandwidth adversary* who controls about 1%. The actual bandwidths being used by the adversary vary over time, as shown by the blue lines in Fig. 2. For both models, when the adversary gets into a guard set, he reduces his relay's bandwidth as long as the total bandwidth of the set remains above 20 MBps. In the low-bandwidth adversary model, the adversary leaves a set if he is the set's main resource provider, which we define as providing more than 90% of the set's bandwidth. In compromising the guard sets, the adversary needs to have only one guard in the set to compromise all the clients attached to that set. If more than one adversary guard is assigned to a set, the adversary will pull one of his guards from that set to inject it into the other set.

We assume that the attacker knows exactly when guard relays will break. In reality, however, the attacker will spend additional resources waiting for guard sets that are close to 20MBps to break. Also, the attacker may miss some guard sets that have sudden large drops in bandwidth. Thus, the results of our experiment represent an upper bound of compromised sets for a given amount of attacker resources.

### 3.2.2 Results

Figure 2 shows the upper bound of the fraction of compromised sets and the ratio of the adversary's bandwidth to the total guard bandwidth in Tor for the first half of 2013. As shown in Figure 2a, a high-bandwidth adversary who owns around 25% of the total guard bandwidth in the Tor network can compromise more than 90% of all the guard sets (and thus be a guard for over 90% of all the clients) in just four months. Figure 2b shows that a low bandwidth adversary with only 1% of total guard bandwidth can compromise around 40% of all the guard sets in just four months. Large fluctuations in the graph are due to large drops in guard bandwidth that occurred in early March and mid-April. Our attacks shows that against the prior guard set design, an adversary with modest resources can endanger the security of a large number of users.

## 4 Design

To mitigate the threat of an adversary compromising a significant number of guard sets, we seek an approach that is more resilient in the face of frequently changing guard sets. In particular, the method should make it harder for an attacker to join a targeted guard set in need of repair and allow clients to keep as much of their anonymity sets as possible, even when guard sets break. To this end, we propose to take advantage of the relative stability of the underlying Internet topology by linking sets to customer cones. In this section, we first explain our motivation for the design, and then we describe how guards are grouped into sets using customer cones and how guard sets are assigned to clients.

### 4.1 Motivation

To prevent the attacks we describe against the guard set design of Hayes and Danezis in the previous section, we need to prevent an attacker from easily joining arbitrary guard sets. First, as Hayes and Danezis also argue [18], we should maintain a hierarchy of guard sets, represented as a tree. This hierarchy dictates that when guard sets are deleted, there is a pre-defined backup guard set for the users of the old set to join. This keeps users together as much as possible, maintaining their anonymity sets. Beyond the Hayes and Danezis proposal, we also would have guards remain in the same place in the hier-archy as much as possible. When a guard set is deleted, the remaining guards should stay in the same general area in the hierarchy, i.e. with siblings in the tree. Also, new guard sets should only be constructed from guards in the same subtree. This prevents guards from attempting to move from one part of the tree to another.

The other major requirement of our approach is that the attacker must not be able to place new nodes into arbitrary locations in the hierarchy. A simple approach would be to use a cryptographic hash of the node's IP address as an identifier, much like in a DHT. Unfortunately, an attacker with even a fairly small range of IP addresses to use could pick a number of different locations in the hierarchy by computing their hash values in advance. If the directory server were to pick the locations of new guards randomly, the attacker could add and remove nodes until the location suited his needs.

In our design, a guard's place in the hierarchy and guard set assignment is based on the guard's network location, meaning the AS it is in and that AS's corresponding place in the customer cones of the Internet. For an adversary with high bandwidth capacity and a range of IP addresses, but only a few network locations, this would substantially limit the number of guard sets he can join and the number of users that he can compromise. To fully overcome this, the attacker would need to be able to place guards into *arbitrary* network locations that have guards. We argue that this attacker model is unlikely in practice. A botnet-based attacker, for example, will likely face challenges with the stability and bandwidth requirements for guards. Even if enough stable bots can be found, the bot locations (such as consumer ISPs) may not correlate well with the locations of Tor guard nodes (which include professional hosting services like OVH), and this further limits the guard sets he can join. We discuss about the hosting providers and their impact on our design in Appendix A.

**Overview.** In the rest of this section, we describe our proposed hierarchy. The hierarchy consists of three levels: 1) *Supersets*, 2) *Sets*, and 3) *Subsets*. A *superset* is a customer cone of a root AS that contains guard ASes. *Supersets* are broken into *sets* and then further into *subsets*. *Sets* represent smaller customer cones within the *superset*, in which all guard ASes have the same provider. Finally, *subsets* are formed by selecting all guards within a *set* and grouping the guards such that the number of ASes within a *subset* are minimized and the guard bandwidth is above a threshold. After this process, each *subset* is then considered a guard set. Below, we describe each part of the system in detail.

## 4.2 *Supersets*

To form *supersets*, we first build a *superset list* – a list of ASes sorted based on customer cone size in ascending order. Initially, the *superset list* contains all ASes with one or more guards (*guard ASes*), and each guard AS is considered as a *superset*. Consider Figure 1 as an example for this section, which means that the list would be something like {AS13, AS34, AS5, AS6, AS7, AS9, AS10, AS11, AS12}, assuming that only the leaf ASes have guards. Then we choose the *superset* with the smallest customer cone size, say AS13. We follow all c2p links to discover all providers for this *superset* that are also providers to at least one other *superset* in the list, e.g. AS8, AS3, and AS1. Among these providers, we select the provider with the smallest customer cone size, e.g. AS8. This provider becomes the new *superset* and is added to the *superset list*, while the *supersets* that are in the customer cone of this provider (AS13 and AS34) are removed from the *superset list*. This process is repeated until all *supersets* in the *superset list* contain guard bandwidths more than a bandwidth threshold $\tau_{up}$ or the number of *supersets* in the list have decreased below a threshold $N$. The pseudo-code of this process is shown in Appendix G.

**Updating *supersets*** As new guard ASes join the network, the algorithm first checks whether the new guard ASes are in the customer cone of an existing *superset*. If they are in an existing *superset*'s customer cone, they are added to that *superset*. If there are still some guard ASes that are not in any of the *supersets*' customer cones, they themselves are considered as *supersets*. Then the above algorithm is run to reform the *supersets*.

## 4.3 *Sets*

*Supersets* often represent large customer cones and many guards. To better isolate groups of guards from each other and make it harder for a malicious guard to move into targeted guard sets, we break each *superset* into *sets*. In building *sets*, the goal is to place guard ASes that are close together in the AS relationship graph into the same *set*. To this end, we first identify all customer cones within the *superset*'s customer cone in which the guard bandwidth reaches the threshold $\tau_{up}$.[2] Note that some cones will be contained within other, larger cones, and there can be overlaps between cones. Among all the possible customer cones, we should pick cones such that their intersection with respect to guard ASes is empty ($A \cap B = \emptyset$). There may be many possible combinations of customer cones that are independent in this way. Since our goal is to have smaller customer cones to make it harder for an attacker to join a targeted *set*, we pick the combination that has the maximum number of independent cones. Each of these independent cones will be a *set* within the given *superset*. At the end, we place all guard ASes that do not meet the requirements for building *sets* into a set.

---

[2] This is the same threshold as used to make the supersets.

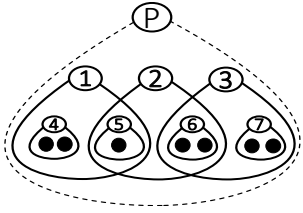

**Fig. 3.** *Set* **creation.** The dashed line shows *superset* P's customer cone, the black circles are guard ASes, and the solid lines show the customer cones with bandwidth $\tau_{up}$ or greater.

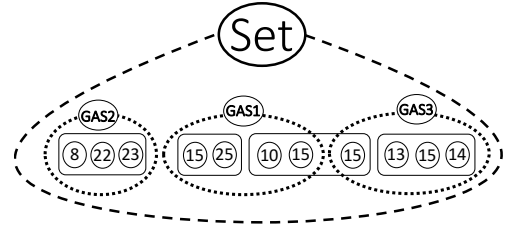

**Fig. 4.** *Subset* **creation.** Dashed ovals represent guard ASes (GAS), rectangles represent *subsets*, and circles represent guards. Numbers inside the circles are the guards' bandwidths (MBps).

Figure 3 shows an example of *set* creation for *superset* P. There are seven customer cones with sufficient bandwidth, but there are overlaps between some of them. The possible combinations of independent cones are {1, 3}, {3, 4, 5}, {2, 4, 7}, {1, 6, 7}, and {4, 5, 6, 7}. Among these combinations, the algorithm picks {4, 5, 6, 7} because it has the maximum number of independent cones. This means that *superset* P has four *set*s.

**Updating *sets*.** Guard bandwidth fluctuates over time, causing some guard ASes to be dropped from the *set* and others added, which requires periodic updates. Our algorithm to update the *sets* first checks which of the new guard ASes are in our current *sets*' customer cones. If a *set*'s bandwidth is below threshold $\tau_{down}$, it dismantles the *set* and releases its guard ASes. The algorithm is then run again to build *sets* from previous *sets*, new guard ASes, and released guard ASes.

### 4.4 *Subsets*

Once we have *sets*, we can break them up further into *subsets*, which are the guard sets themselves. We first randomly shuffle the guard ASes in the *set*. Then we add one guard at a time from the same AS to the current *subset* until the *subset*'s bandwidth reaches the threshold $\tau_{up}$.[3] If we use all the guards in an AS, we continue adding guards from the next guard AS.

Figure 4 shows an example of *subset* creation. The *set*'s customer cone includes three guard ASes, GAS1, GAS2, and GAS3, and four *subsets* are formed. Note that a *subset* can include all of the guards in an AS (such as in GAS2), some of the guards in an AS (such as the leftmost *subset* in GAS1), or guards from multiple ASes (the two rightmost *subsets*).

**Updating *subsets*** *Subsets* will need to be updated over time due to the leaving and joining of guard ASes and changes in bandwidth. If a *subset*'s bandwidth falls below the threshold, $\tau_{down}$[4], we need to repair that *subset* to ensure load balancing. To repair a low-bandwidth *subset*, we add new guards to it until the bandwidth reaches at least 40 MBps. We first try to add new guards that are in the same AS, and then add from other guard ASes in the *set*. If there are still some unused new guards in the *set*, we try to build new *subsets* out of them.

### 4.5 Assigning Clients to the Guard Sets

A newly-joined client selects first a *superset*, then a *set* from among the *sets* in her *superset*, and finally a *subset* from among the *subsets* in her *set*. Each of these selections is random, weighted proportionally by bandwidth. To create a circuit, the client picks one of the guards in her guard set as the entry relay. The selection of guards from a guard set can be weighted in favor of bandwidth or can be uniformly at random.

As time passes, some guards leave the network, and this causes some *subsets*, *sets*, or even *supersets* to be no longer available. If the client's guard set has been dismantled, the client will select another *subset* under her *set*. Similar recovery methods are available for *sets* and *supersets*. In the worst case, if her *superset* is gone, the client acts like a newly-joint client. The pseudocode for these processes is shown in Appendix G.

---

[3] Again, this is the same threshold as used for *supersets*.
[4] The same threshold as for updating *sets*

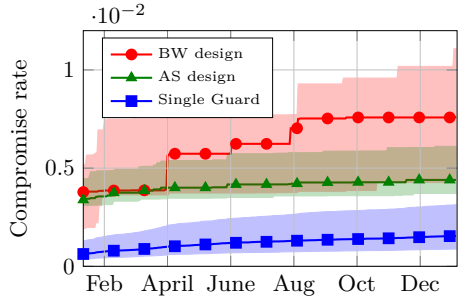

**Fig. 5.** *Low-resource* adversary: the solid lines show the median of compromised clients and the colored bands shows the area between the first and third quantiles.

# 5 Evaluations

In this section we evaluate different aspects of our guard set design. We start by analyzing the security of the proposed guard set design in the presence of relay- and network-level adversaries. Then we monitor guard set changes over time with respect to the number of guard sets, guard set bandwidth, and client anonymity sets. In our evaluation, we set bandwidth thresholds $\tau_{up} = 40$ MBps and $\tau_{down} = 20$ MBps – the same as the thresholds used by Hayes and Danezis [18]. We set the number of supersets to $N = 50$ to maintain enough *Supersets* and not to be merged into only top tier ASes. For our data set, we use consensus documents from January 2015 to January 2015 from Tor Metrics [36]. Following Hayes and Danezis, during the entire evaluation we do not rotate the guards. Additionally, we compare our results to Hayes and Danezis's design [18], which we refer to as "*BW* design." We call our design "*AS* design."

## 5.1 Security Evaluation

Tor is known to be vulnerable to traffic correlation attacks [19, 24, 26, 28, 35]. An adversary who observes both entry and exit traffic can use the timing of packets to link clients to their destinations. Observing both sides of Tor traffic can happen at the relay level or the network level. At the relay level, an adversary running guard and exit nodes in the Tor network may deanonymize clients whose circuits traverse the adversary's guard and exit relays. We examine two relay-level attack scenarios, a non-targeted relay-level adversary and a targeted relay-level adversary. At the network level, the adversary controls some part of the network, such as an Autonomous System (AS) or Internet Exchange Point (IXP), and can thus observe huge amounts of traffic, including entry and exit traffic in Tor. In this section, we examine the security of our guard set design against both relay-level and network-level adversaries.

**Security Claims.** In this section, we seek to demonstrate the following:

1. The compromise rate is lower for *AS* design compared to *BW* design for a variety of relay-level adversaries with varying resource levels and for both non-targeted and targeted attacks.
2. The vulnerable stream rate is approximately the same as both Tor and *BW* design against network-level adversaries.
3. *AS* design is compatible with the DeNASA [9] AS-aware path selection algorithm, and the combined algorithms provide similar vulnerable stream rates as DeNASA against network-level adversaries.

### 5.1.1 Non-Targeted Relay-Level Adversaries

A non-targeted relay-level adversary adds guard nodes in the network to compromise guard sets. This adversary does not target any specific guard set or client; his goal is to compromise as many clients as possible, which means joining as many guard sets as possible.

**The adversary model.** If a guard set contains one compromised guard relay, we consider the entire guard set to be compromised; all clients using that guard set will be compromised because they eventually send traffic through the compromised guard. This follows the model of Hayes and Danezis [18]. We examine the relationship between the amount of guard bandwidth the attacker provides to the Tor network and his success rate in compromising guard sets. For each of the guard selection strategies, *AS* design, *BW* design, and *Single Guard* (i.e. Tor), we assume that the adversary runs some guard relays such that their total bandwidth adds up to 1%, 5%, or 10% of the total guard bandwidth of Tor for different experiments.

Because *AS* design uses both bandwidth and AS relationships, the adversary's network (AS) matters as well. Therefore, we analyze the security of our guard set design under three attack strategies: a *low-resource* adversary, a *high-resource centralized* adversary, and a *botnet* adversary. In Tor, a client is considered compromised if it chooses a malicious guard. We note that this is not completely fair to the guard set designs, since a

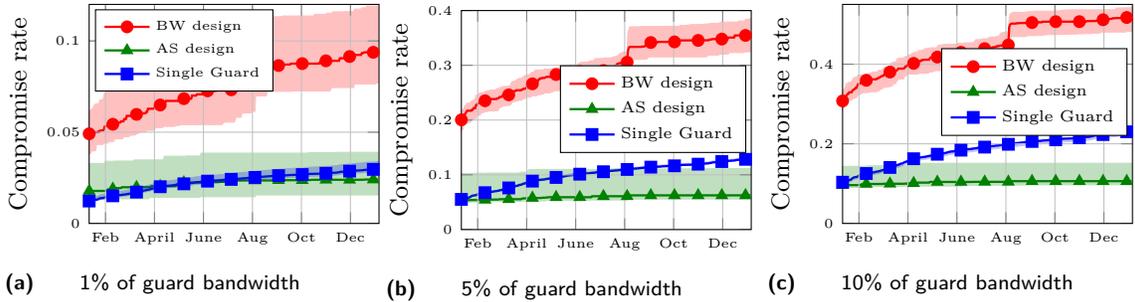

**(a)** 1% of guard bandwidth  **(b)** 5% of guard bandwidth  **(c)** 10% of guard bandwidth single guard in Tor can compromise more of the user's traffic than any one member of a guard set in which each guard is picked only part of the time. We allow the adversary to inject malicious relays at the beginning of the simulation before the 500,000 users that we simulate start picking their guards. Unlike the model used by Hayes and Danezis [18], we assume that the adversary's guard relays remain up and available during the entire simulation, which gives the adversary an advantage.

**Fig. 6. High-resource centralized adversary.** Compromise rates for varying guard bandwidth controlled by the adversary. The solid lines show the median of compromised clients and the colored bands show the area between the first and third quantiles.

**Low-resource adversary.** In this attack strategy, we assume that the adversary injects only a single guard relay in the network. We randomly choose an AS for this malicious guard and select its IP address randomly from the IP range of the selected AS. We also randomly select a bandwidth value from Tor guards' bandwidths in the consensus document. This malicious guard is added to the network, and the simulation is run for the year of consensus files. We repeat the simulation 50 times, with a new malicious guard each time.

Figure 5 shows the median fraction of compromised clients over 50 simulations. At the beginning of 2015, the compromise rate of *AS* design is statistically similar to *BW* design. The compromise rates of both *AS* design and *BW* design are greater than *Single Guard*, because it is assumed that a single malicious guard within a guard set compromises all clients who choose that guard set.

Over time, the compromise rate in *BW* design grows substantially from 0.036% to 0.076%, as the malicious guard moves into different guard sets. On the other other hand, *AS* design's compromise rate only grows from 0.032% to 0.044%. We note that the variance in these results for low-resource adversaries is high, as seen by the wide quartile bands, but the trends are consistent. The growth of the compromise rate in *AS* design is 37%, much smaller than the 110% growth in *BW* design. The reason for this is that in *AS* design, malicious guards are quarantined inside a *set* within a *superset*.

By design, the malicious guard cannot infiltrate guard sets that are outside of the malicious guard's *set*. In contrast, in *BW* design, when the bandwidth in the network changes or the malicious guard's bandwidth changes, the guard sets change and the malicious guard moves from one guard set to another and compromises additional guard sets over time.

**High-resource centralized adversary.** In our model, the high-bandwidth adversary owns some relays such that its total bandwidth is a considerable fraction of the network's total bandwidth. We assume that the adversary is *centralized*, meaning that it injects all malicious guards into a single AS. We select at random one guard AS in which to add the malicious guard relays, and we select their bandwidths randomly from live Tor guard bandwidths. Such relays are added to the Tor network until the target attacker bandwidth is reached. The simulation is run 50 times, with a new guard AS and new malicious relays each time.

Figure 6 shows the fraction of compromised clients for three different guard selection schemes and three different adversarial bandwidth assumptions. As the adversary's bandwidth increases, the fraction of compromised clients increases for all three schemes. *BW* design has the largest fraction of compromised clients compared to the other two schemes. In *BW* design, we observed a significant increase in compromises in August 2015. We found that this was due to the large drop in Tor guard bandwidth in August 2015. The daily guard bandwidth in the Tor network, as shown in Figure 17 in Appendix F, significantly decreased in August for a few days. This triggered a significant churn in guard sets for *BW* design that allowed malicious guards to infiltrate more guard sets. This suggests that bandwidth changes in the Tor network have a negative impact on security for *BW* design due to its vulnerabilities.

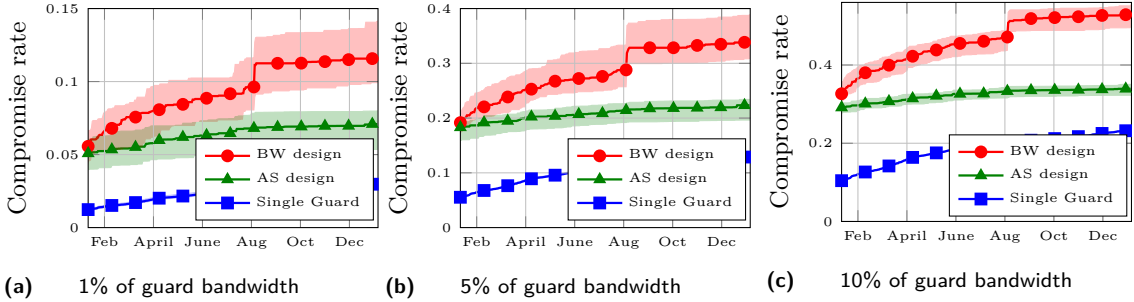

**Fig. 7. Botnet adversary.** Compromise rates for varying fractions of total guard bandwidth owned by the adversary.

After one year of simulation time, we observed significantly higher growth in the fraction of compromised clients for *Single Guard* compared to *AS* design for all three adversarial models. For 1%, 5%, and 10% simulated adversarial bandwidth, the fraction of compromised clients for *Single Guard* increased by 180%, 168%, and 157% over one year, respectively. The fraction of compromised clients for *AS* design remained almost constant over one year. These results support our assertion that *AS* design successfully constrains the adversary's guards to guard sets within the *set* and *superset*. Moreover, clients do not rotate guard sets unless their guard set is broken up. Even then, the client will choose another guard set within its *set* and *superset*. These characteristics allow the *AS* design to keep the compromise rate low over time.

We also explore how our results change for different bandwidth thresholds $\tau_{up}$ and $\tau_{down}$ in both *AS* design and *BW* design. Figure 8 shows the compromise rates for three different values of $\tau_{up}$, while keeping $\tau_{down} = \tau_{up}/2$. The compromise rate of *BW* design increases significantly as $\tau_{up}$ increases. On the last day of simulation, for example, the compromise rate for $\tau_{up} = 60MBps$ ($\tau_{down} = 30MBps$) is 0.42, a 44% increase over that of $\tau_{up} = 30MBps$ (0.29). The guard sets in *BW* design are more prone to breaking for larger values of $\tau_{up}$ and $\tau_{down}$, which causes more guard rotation and an increased rate of compromise. For *AS* design, we observe only small changes in the compromise rate for changing bandwidth thresholds. For $\tau_{up} = 60MBps$ ($\tau_{down} = 30MBps$), the compromise rate goes from 0.056 on the first day to 0.068 on the last day of simulation, where the latter is only a 13% increase over the compromise rate when $\tau_{up} = 30MBps$ (0.06). This is due to the use of network location rather than bandwidth as the key criteria for managing sets in *AS* design. Note that for clarity, due to how close the results are, Figure 8 only shows the results for $\tau_{up} = 40MBps$.

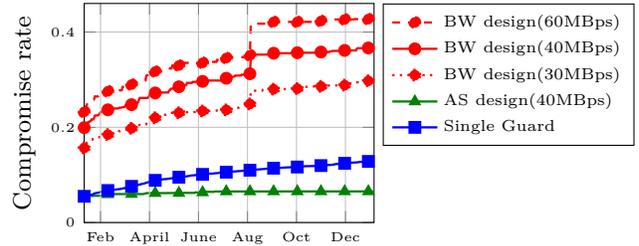

**Fig. 8. High-resource centralized adversary.** Compromise rates as $\tau_{up}$ varies, $\tau_{down} = \tau_{up}/2$, and the adversary controls 5% of guard bandwidth.

**Botnet adversary.** This adversary is similar to the high-resource centralized adversary, except the adversary injects his guard relays from different guard ASes instead of one guard AS. For each simulation, the adversary adds malicious guards to the network from different guard Ases, selected randomly from all guard ASes, until the desired bandwidth for adversary is reached. We also repeat this simulation 50 times.

Figure 7 shows the fraction of compromised clients for all three guard selection schemes in the presence of a *botnet* adversary with different bandwidth fractions. We see that the results for BW design and *Single Guard* do not change compared to the *high-resource centralized* adversary (Figure 6) because these two methods work with bandwidth in either grouping guards or client assignments and do not use the guard ASes. Because the adversary's relays are in different ASes, however, he can compromise many more guard sets in the AS design. As we see in the figures, *AS* design's compromise rate in this attack is higher than its compromise rate in Figure 6. Nevertheless, *BW* design's compromise rate rises much faster than *AS* design's, which is almost constant over time. This indicates that, although this adversary can compromise many sets, the AS design is good at stopping the propagation of the adversary's impact on the network. *Single Guard* has a compromise rate less that the other two guard set designs, which is a trade-off against the smaller anonymity sets provided by *Single Guard*. The growth of compromise rate over a year against *Single Guard* is 194%, 168%, and 155% for 1%, 5%, and 10% simulated adversarial bandwidth, respectively. These compromise rate growths are much higher than 62%, 17% and 22% growth seen against *AS* design for 1%, 5%, and 10% simulated adversarial bandwidth, respectively.

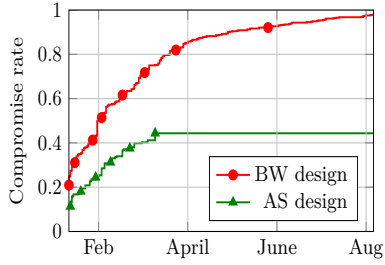

**Fig. 9. Targeted attack.** CDF of time to compromise

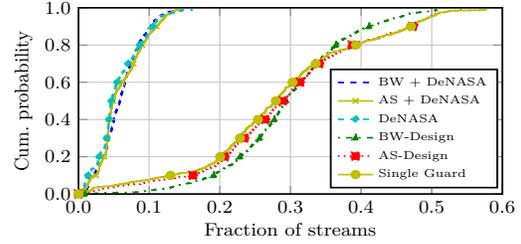

**Fig. 10. AS level adversary.** CDF of vulnerable stream rates

#### 5.1.2 Targeted Attacks

The other case we examine is when the adversary targets a specific client, adding guard relays to Tor with the goal of getting added to the client's guard set. The way *AS* design builds and repairs the sets makes it harder for the adversary to get into the targeted client's guard set.

**The adversary model.** The attacker controls one or more exit nodes and profiles a particular user of interest based on her exit traffic. We assume that this adversary can identify the client's guard set. Given that all clients know the assignment of guards to guard sets, an adversary simply needs to run one Tor client to get this mapping. It is more challenging to learn the assignment of the client of interest to her guard set. A variety of attacks, however, reveal the clients' guard relays [16, 19, 30, 32, 33], and we assume that the attacker uses one of these attacks successfully to identify one of the guards and the corresponding guard set. Then, the goal of the attacker is to get into this guard set.

We simulated the attack for *BW* design and *AS* design. In the simulations, we select a target client and assign a guard set to this target on the first day. Then the adversary monitors the network, measures the guard bandwidths, and waits until the target's guard set is about to break. We assume that the attacker can add some guard relays whenever he chooses—we discuss the validity of this assumption in Section 3.

Against *AS* design, when the target guard set is broken, the attacker selects one of the guard ASes used in that guard set as the AS from which he injects his guard relays. The attacker finds all the broken guard sets in which this AS belongs and computes the total amount of bandwidth needed to fix all of those broken guard sets. Then the attacker adds guard relays from the chosen AS to the network until they reach the required bandwidth. The attack against *BW* design proceeds as described in Section 3.

We simulated the targeted attack for the course of one year from January 2015 to December 2015. On the first day, January 1, we added a target client and assigned a guard set to her based on the given guard set assignment algorithm. Once the target guard set is determined, we wait until the target guard set is broken. We then add the adversarial relays to the network and tune their bandwidth based on the bandwidth needed to get into that set. We continue monitoring the network, tuning the adversary's bandwidth and adding more guard relays as needed until the target is compromised. When the target is compromised, the adversary tunes his bandwidth to keep the guard set alive, i.e. the target set's bandwidth should not be less than $\tau_{down}$. We repeat this process for 500 targeted clients.

**Results.** Figure 9 shows the cumulative fraction of days it takes for the targeted clients to be compromised. With *BW* design, almost all the targeted clients (98%) were compromised within the year, and 50% of the targets were compromised within one month. *AS* design protects clients better from the targeted attack, as just 44% of targets were compromised within the year. The reason for this is that if the target guard set is broken, it is repaired by guard relays from ASes in the same *set*. This limits the chances for the adversarial AS to be picked to repair the set. Overall, we find that clients in *AS* design are safer than in *BW* design.

### 5.1.3 AS-Level Adversaries

An adversary may be able to monitor network traffic on one or more ASes or IXPs on the Internet and observe both sides of Tor circuits to link users with their destinations. In this study, we consider a stream to be *vulnerable* if both the entry and exit sides of the traffic traverse the same AS. To examine the security of our design at the AS level, we implemented the guard selection schemes in TorPS [24] and generated streams from a set of clients to a set of destinations. Then we found the AS paths on both the forward and reverse connections from client to guard, and exit to destination [35]. We had 6,000 clients connecting from 30 client ASes distributed over the top countries using Tor. To choose the 30 ASes, we first pick a country for a user based on the distribution of users from the top countries according to Tor Metrics [7]. We then check whether this country has an AS in top client ASes list given by Edman et al [14]. If so, we pick that AS and add it to our client AS list, and remove that AS from the top client ASes list. Otherwise, we randomly select an AS from that country and add it to our client AS list. We keep selecting countries based on the distribution of directly connected clients distribution until we have 30 client ASes. Over one month of simulation time (Feb. 2015), these clients generated 8 million streams for each guard selection mechanism.

Figure 10 shows the CDF of vulnerable stream rates for clients using *AS* design, *BW* design, and *Single Guard*. As shown, the fraction of vulnerable streams is almost the same for the three guard selection mechanisms with a median of 28%. Thus, *AS* design appears to provide similar anonymity as *Single Guard* and *BW* design against an AS-level adversaries.

We combined the guard selection schemes with DeNASA [9], an AS-aware path selection algorithm. DeNASA avoids paths with *suspect* ASes, mainly Tier 1 ASes that appear frequently on the entry and exit sides of Tor traffic. We provide a brief overview of DeNASA in Appendix E. As shown in Figure 10, after combining the guard selection mechanisms with DeNASA, the median vulnerable stream rate for both guard set designs and *Single Guard* dropped 80% (from 0.28 to 0.05). Thus, we believe that *AS* design is compatible with DeNASA for protecting against AS-level adversaries.

## 5.2 System Evaluation

We now examine dynamics in the number of sets, sets' bandwidths, and anonymity sets.

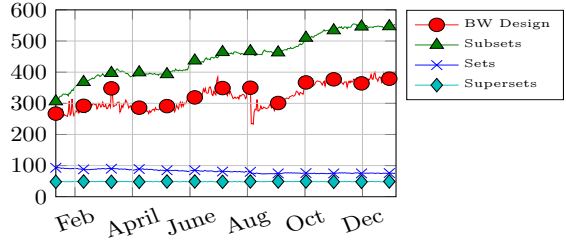

**Fig. 11.** Counts of *AS supersets*, *sets*, and *subsets* and *BW* sets

**System Claims.** In this section, we seek to demonstrate the following:
1. The hierarchy derived from the customer cones is stable over time, with moderate changes to sets and subsets.
2. The anonymity sets of users are significantly higher for *AS* design over *Single Guard* and approximately the same as *BW* design.
3. Bandwidth is distributed sufficiently evenly between guard sets in *AS* design to not create bottlenecks at the guard or waste significant bandwidth.
4. Network performance in Tor is approximately the same for *AS* design compared with *Single Guard*.

**Guard Sets.** Our guard set design includes *supersets*, *sets*, and *subsets*. Figure 11 shows the number of each of these elements over the year 2015. The number of *supersets* was 48 and did not change throughout the experiment. The number of *sets* changed modestly over time, ranging from 76 to 92 with an average of 84. This shows that the customer cones that make up the *sets* do not change greatly over time. *Subsets* increased over time because they are built at the relay level and based on relay bandwidth. Figure 17 shows the daily guard bandwidth in the Tor network, and we observe three significant increases in bandwidth: in late February, late June, and early October. These increases in bandwidth correspond closely with the increase of *subsets* in Figure 11. During our simulations, on average 15 guard sets in *AS* design got repaired each day, compared with seven guard sets in *BW* design. *AS* design often adds guards from broken sets to existing sets, and this counts as a "repair" of the existing set.

**Anonymity Sets.** We define an anonymity set as a set of clients that use the same set of guards. If the size of the anonymity set is small, then the threat of the guard fingerprinting and statistical disclosure attacks increase. To evaluate the anonymity set sizes in our design, we use our client assignment mechanism described in Section 4.5 to attach clients to guard sets.

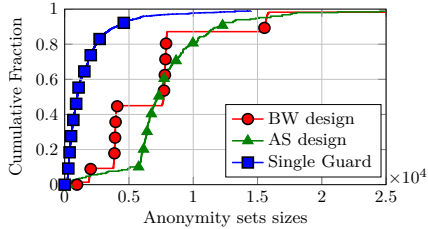

**Fig. 12.** CDF of anonymity set sizes

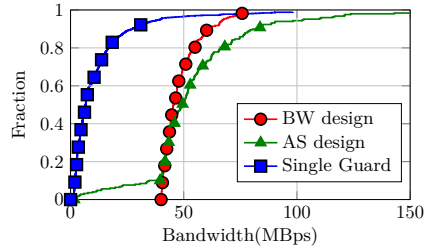

**Fig. 14.** CDF of guard sets' bandwidths

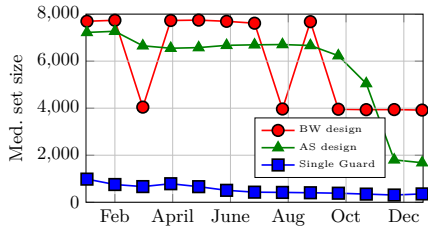

**Fig. 13.** Median of anonymity set sizes over time

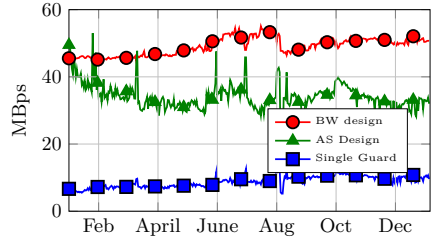

**Fig. 15.** Bandwidths of guard sets over time

In this experiment, we model 2,000,000 clients, which is approximately the number of Tor daily users [36].

We compare the anonymity set sizes in *AS* design with the ones in *BW* design and Tor (*Single Guard*). Figure 12 shows the empirical CDF of anonymity set sizes. The median anonymity set size in *AS* design (7300 clients) is almost the same as *BW* design (7700 clients), and both are far larger than in Tor currently (970 clients). Figure 13 depicts the changes in median anonymity set size over time. The median for both *BW* design and *AS* design decrease over time. Because we do not rotate guards and do not consider user churn in our simulations, new guard sets have few users and small anonymity set sizes. The sizes decrease particularly fast from September, which corresponds to when the number of guard ASes starts increasing most rapidly.

**Set Bandwidth.** To ensure network performance remains similar to Tor, we must ensure that guard bandwidth is distributed relatively evenly over guard sets. To do this, we test whether all guard sets have an accumulative bandwidth above a certain threshold.

Figure 14 shows the CDF of bandwidths among the guard sets in *AS* design and *BW* design and among individual guard relays in *Single Guard*. The results for *AS* design and *BW* design are similar, though *AS* design has more high-bandwidth guard sets. This is because it does not break up a high-bandwidth guard into multiple sets the way that *BW* design does. Note that *AS* design retains good load balancing in this case by having clients pick their guard sets with a weight for bandwidth.

Figure 15 shows the median guard set bandwidth over time in *AS* design and *BW* design and the median guard relay bandwidth for *Single Guard*. The average of the median guard set bandwidths over the year 2015 are 34 MBps and 49 MBps for *AS* design and *BW* design, respectively. Although we used the same bandwidth thresholds $\tau_{up}$ and $\tau_{down}$ as Hayes and Danezis [18], *BW* design's guard sets tend to have more bandwidth because bandwidth is the primary consideration for managing guard sets. In *AS* design, network location is the primary consideration and bandwidth is secondary.

**Performance.** To evaluate the performance of the guard selection mechanisms, we simulate them using Shadow [5, 22], a discrete-event network simulator that runs real applications like Tor and BitCoin on a single machine. Using Shadow, we simulate *Single Guard*, *AS* design, and *BW* design on a Tor network with 742 relays (including 152 guard relays) and 2700 clients (including 2280 web clients). Details about the Shadow configuration and simulation setup can be found in Appendix C.

The median of the times to download the first byte (TTFB) for clients are 0.55, 0.55, and 0.56 seconds in *AS* design, *Single Guard*, and *BW* design, respectively. Thus, the responsiveness is about the same in all three approaches. The median times to download the last byte for web clients (TTLB) are 1.221, 1.185, and 1.183 seconds in *BW* design, *AS* design, and *Single Guard*, respectively. Here, all the mechanisms have almost the same throughput.

# 6 Discussion

In this section, we discuss the implications of our findings and the scope for future work. Further discussion about the guard rotation can be found in Appendix D.

**Deployment.** To implement guard sets in Tor, the Tor directory authorities would be responsible for building and disseminating the guard sets. In particular, they would use information on p2c links from CAIDA [3] to identify customer cones and build the tree structure of *AS* design, including *supersets*, *sets*, and *subsets*. They then ship the information about the design structure such as list of *supersets*, *sets*, *subsets*, and which guard relays are in the *subsets* to the clients using the consensus documents. The clients will get the information about the guard sets through the consensus documents. Then clients only need to pick their guard set using the algorithm mentioned in Section 4.5. We propose a format for adding guard set information to the documents in Appendix B. We used this format to add the guard set information to consensus documents for 2015. We observed that the average document size was increased grew 3% (from 1.507 to 1.550 MB).

Since not all Tor clients will upgrade at once, some degree of incremental deployment is needed. One approach is to have the directory servers provide just the existing consensus documents to older clients and the additional guard set information to upgraded clients. To prevent a major partitioning of clients into anonymity sets based on different behavior, Tor can wait until many clients have upgraded before setting a flag that initiates the use of guard sets. It may be best if transition to guard sets happens done slowly, as users could continue with their current guards and not rotate from them prematurely. Beyond this, further studies should be conducted on the topic of incremental deployment to understand the impact of different transition strategies on the anonymity of users and performance of the system.

**Orphan guard ASes.** There are some cases in which guard ASes are isolated in a customer cone by themselves. In such cases, our algorithm may not be able to group these guard ASes with other guard ASes, forcing them to form smaller guard sets. If these sets do not offer enough bandwidth, they pose a risk for guard fingerprinting. To mitigate this issue, we can ignore these low bandwidth guard sets until enough guard relays join the network such that the low bandwidth guard sets will be merged to form a sufficiently high bandwidth guard set. In our one-year simulations, we observed a median of two such ASes and a maximum of six ASes, and the median bandwidth of these ASes were 3.6 MBps.

**AS relationships and customer cones.** Customer cones are not as simple in practice as the tree model would suggest. Given the CAIDA AS relationship graph, which itself is not completely accurate, the cones must be inferred based on some assumptions. For our work, we apply the recursive customer algorithm [12, 27] (see §2.1). Other approaches might yield different results, but the key feature we need is that the attacker cannot add guards to arbitrary points in the hierarchy. Issues such as accuracy of the cones and the presence of p2p traffic passing through IXPs should not affect our findings much because they affect the AS graph but not the relative stability and hierarchical nature of the AS graph that is being exploited by the proposed technique.

We note that relying on a single organization such as CAIDA to form customer cones may be vulnerable to attacks on the organization or its information. Further thinking and experimentation should be performed before deploying these techniques in Tor.

To evaluate the risk of AS-level adversaries, we used Qiu and Gao AS-level path inference [34]. Although this type of inference can be inaccurate in identifying all of the ASes on a path [25], Barton and Wright report that it is 90% accurate in identifying the eight most common ASes that appear on both ends of a Tor path [9]. Further, they find that these eight ASes account for about 98% of all instances of an AS appearing on both ends of a Tor path. Thus, our findings both with and without DeNASA should provide a reasonably accurate estimate of the risk of network-level attacks.

Finally, we note that IP-to-AS mapping is not perfect. BGP is not secure, which can be leveraged for attacks on Tor [35]. Thus, it can be similarly attacked to undermine the *AS* design, and the importance of this requires further investigation.

**Virtual Hosting.** *AS* design takes the advantage of this fact that the placement of guard relays in a given AS is harder than manipulating the bandwidth, which is the approach offered by *BW* design. There are some organizations and hosting providers like OVH that contribute a significant fraction of guard relays to the Tor network. Table 1 in Appendix A shows the top 20 organizations running guard relays. The adversary can run a guard relay on hosting provider like OVH and and get into the guard sets built based on OVH's AS. This allows the adversary to target those guard sets beyond what we have discussed in this paper. On the other hand, for a large provider like OVH, there will be multiple guard sets, and no single one can be targeted. Further, the adversary does remain confined to his *set* and *superset*. Closer examination of this issue is needed before $AS$ design can be deployed in Tor.

**DoS Within Sets.** In $AS$ design, if the client's guard set is dismantled, the client will select another subset from her set. This confines the adversary better to her *set* and limits the number of clients the attacker can compromise. On the other hand, the adversary can DoS other guard sets in the same *set* and force their clients to pick her guard set. This can be limited by picking the new guard set from a larger region in the network, such as anywhere in the *superset*. Since this creates more opportunities for the adversary to compromise guard sets, and since DoS attacks are active and detectable, we argue that it is better to select from the set instead.

# A Hosting Providers

| organization | No. of relays | No. of ASes | VPS |
|---|---|---|---|
| Digital Ocean, Inc. | 63 | 7 | ✓ |
| LeaseWeb Netherlands | 18 | 2 | ✗ |
| ITL Company | 13 | 2 | ✗ |
| SoftLayer Tech. Inc. | 12 | 2 | ✓ |
| Eonix Corporation | 7 | 2 | ✓ |
| PJSC Rostelecom | 5 | 2 | ✗ |
| Not SURF Net | 4 | 2 | ✗ |
| JSC "ER-Telecom" | 2 | 2 | ✗ |
| OVH SAS | 227 | 1 | ✓ |
| ONLINE S.A.S. | 141 | 1 | ✓ |
| Hetzner Online GmbH | 122 | 1 | ✓ |
| myLoc managed IT AG | 28 | 1 | ✓ |
| ISPpro Internet KG | 17 | 1 | ✓ |
| Contabo GmbH | 17 | 1 | ✓ |
| domainfactory GmbH | 16 | 1 | ✓ |
| MCI Com. Services, Inc | 13 | 1 | ✗ |
| Seflow S.N.C. | 13 | 1 | ✓ |
| Strato AG | 12 | 1 | ✓ |
| Init7 (Switzerland) Ltd | 12 | 1 | ✗ |
| PlusServer AG | 12 | 1 | ✓ |

**Table 1.** Top 20 organizations, sorted by the number of ASes, running guard relays in Sept. 1, 2015, we used CAIDA dataset to map the ASes to the organizations [2].

# B Consensus document

As we mentioned in Section 6, the directory authorities can build the guard sets and ship them to the clients through the consensus documents. We propose the following format for adding the guard set's information to the consensus documents. This format is compatible with the formats used in consensus documents explained in [6].

Each router entry in the consensus documents contains a set of items. Each item sits in a separate line which is started with an identifier. To include the guard set information to the consensus documents, we add one more item for each guard relay. This item has the following format:
`"g" SP SUPERSET-ID SP SET-ID SP SUBSET-ID NL`
Where:
`"g"` = the identifier.
`SP` = white space.
`SUPERSET-ID` = 16–digit unique *Superset* identity.
`SET-ID` = 16–digit unique *Set* identity.
`SUBSET-ID` = 16–digit unique *Subset* identity.
`NL` = new line.
We used this format to add the guard set information to consensus documents in the year 2015. We observed that the average document size was increased 43 Kbytes (from 1.507 to 1.550 Mbytes).

# C Shadow Configuration

In our Shadow simulations, we follow Tor modeling procedure suggested by Jansen et al. [20]. We used Tor version 2.4.26 (released in March 2015) and modified it to implement *AS* and *BW* designs. Shadow comes with some validated tools that use the data from Tor Metrics [36] and generates the private Tor network. We found a problem in one of the Shadow's tools which was affecting the computing of the relays load. We fixed the problem and reported it. Using those tools and Tor Metric data from April 2015 (one month after the release of Tor version 2.4.26), we generated a Tor network with 742 relays (including 152 guards, 63 exits, 5 Authorities, 40 exitguards, and 482 middles) 800 HTTP servers, 2700 clients (2280 web clients, 120 bulk clients, and 300 Shadowperf clients).

Shadow runs the actual Tor source code over a simulated Internet topology. The default Internet topology shipped by Shadow is very small, with 183 vertices and 17,000 edges, which makes this topology not a good representative of the Internet. For our simulations, we used the topology used by Jansen et al. [21]. This topology is built by techniques from recent studies in modeling Tor network [20, 24], and the data from Tor Metrics [36], CAIDA [3], and Alexa [1]. This topology contains 699.029 vertices and 1,338,590 edges. Shadow developer Rob Jansen provided us this topology that uses a latency-based packet loss model where the higher latency equals higher packet loss.

Figures 16 shows the performance results for different guard selection mechanisms and compares them with TorPerf clients [4]. As shown in the figure all approaches have the same performance results. The median of the time to download the first byte is almost 0.55 seconds for all the approaches (it is 0.68 seconds in TorPerf clients). The median of the download time for 320KiB objects is around 1.2 seconds for all approaches which is between 1.06 seconds (the median of download time of 50KiB-TorPerf clients) and 3.2 seconds (the median of download time of 1MiB-TorPerf clients). In our simulations, $AS$ design built 58 guard sets and $BW$ design built 50 guard sets, they built 394 and 275 guard sets, respectively, in April 15, 2015, the day we used the data to generate our private Tor network.

## D Further Discussion

**Guard set rotation.** Like Hayes and Danezis [18], we did not include guard set rotation in our system since clients may rotate to malicious guard sets [24]. On the other hand, without guard set rotation, old guard sets will collect more users over time compared to new guard sets. Moreover, we do not let compromised clients regain their privacy by rotating their guard sets. This trade-off also exists in other guard selection schemes. In Tor's single guard selection policy, the client should rotate her guard every 9 to 10 months. This guard rotation period is long enough that guard rotations are significantly reduced, yet compromised clients eventually regain their privacy. Such a guard rotation policy can also be applied to the guard set designs. We can also consider an age metric for guard sets that keeps the number of clients from growing too much in old guard sets.

## E DeNASA Implementation

DeNASA [9] is a recently proposed AS-aware path selection algorithm. It avoids paths with *suspect* ASes, mainly Tier 1 ASes that appear frequently on the entry and exits sides of Tor traffic. The main advantage of this approach is that it is destination-naive, which enables Tor to preemptively build circuits for performance reasons. The downside of DeNASA is that it is vulnerable to leakage about a client's AS across repeated connections [23].

We implemented the *g&e*-select algorithm introduced by DeNASA. In combining DeNASA with guard set designs, the client first picks a guard set (or a guard relay in *Single Guard*) in such a way that is defined in each guard selection mechanism. If there is a suspect AS on the AS path between the client and any of the guards in the chosen guard set, the client drops that set and tries another guard set. Otherwise, she keeps the guard set. At the time of building a circuit, DeNASA picks a guard relay from the chosen guard set; then it picks an exit relay such that Tor picks, if the probability of appearing suspect ASes existing on the entry path (the AS path between the client and and the guard relay) is less than a threshold, the chosen exit is acceptable. Otherwise, it tries another exit relay. We set the probability threshold to 0.1 in our simulations. .The suspect ASes for the entry side, the suspect ASes which appear more frequently on the path between the clients and guard relays, are :{ 1299, 3356 }.

The suspect ASes for the exit side, the suspect ASes which appear more frequently on the path between the exit relays and destinations, are :{ 1299, 3356, 6939, 174, 2914, 3257, 9002, 6453 }.

The probability table is the input to DeNASA algorithm. The rows in the table are exit ASes, which have exit relays, and columns are the suspect ASes. In the table, each value $P_{ij}$ represents the probability of appearing suspect AS $j$ on the AS paths between exit AS $i$ and the possible destinations. We considered the possible destinations all the destinations visited by TorPS *typical* user model.

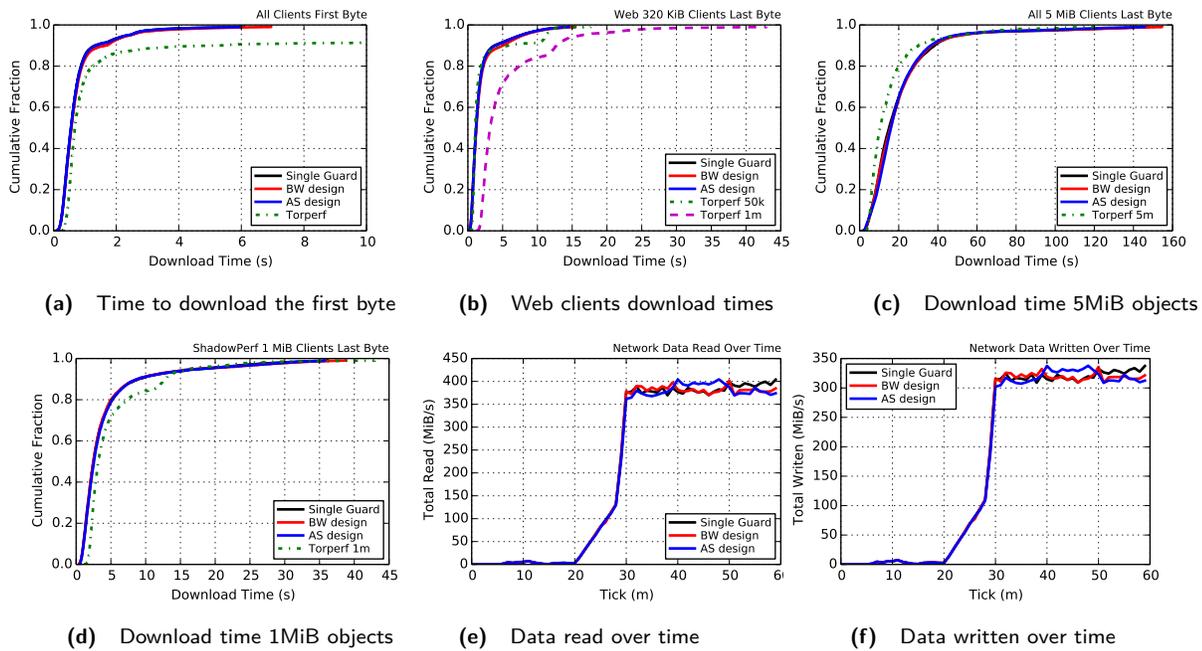

Fig. 16. **Performance Results.** The performance evaluation of guard selection mechanisms.

# F Daily Bandwidth information

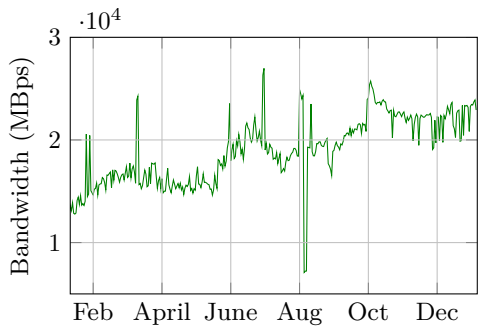

Fig. 17. Daily guard bandwidth throughout 2015

# G Algorithms

```
input  : Guard ASes, customer-provider graph,
         τ_up, N
output: Supersets

%initialization;
superset_list ← Guard ASes;
superset_list ← SORT ascending superset_list based
  on their customer cone size;
while TRUE do
    superset_current ← superset_list.pop(0);
    providers list ← GET superset_current's
      providers from customer-provider graph;
    providers list ← SORT ascending providers list
      based on their customer cone size;
    for provider ∈ providers list do
        for superset ∈ superset_list do
            if superset ∈ provider.cone then
                DELETE superset from
                  superset_list;
                APPEND provider to superset_list;
                BREAK;
            end
        end
    end
    if length(superset_list) < N_superset then
       BREAK
    end
    HIGH BW SETS ← 0;
    for superset ∈ superset_list do
        if superset.bw > τ_up then
            HIGH BW SETS ←
              HIGH BW SETS + 1;
        end
    end
    if HIGH BW SETS == length(superset_list)
     then
       BREAK;
    end
end
```
**Algorithm 1:** The algorithm building *supersets*

```
input  : Supersets customer-provider graph
output: Guard Set

Function Selection(list):
    totalbw ← 0.0;
    rand ∈ random(0, totalbw);
    for item ∈ list do
        totalbw += item.bandwidth;
    end
    rand ← random(0, totalbw);
    tmp ← 0.0;
    for item ∈ list do
        tmp += item.bandwidth;
        if tmp > rand then
            return item;
        end
    end
end

superset ← SELECTION(Supersets);
set ← SELECTION(superset.sets);
guardset ← SELECTION(set.subsets);
return guardset, set, superset;
```
**Algorithm 2:** Picking guard set